\documentstyle[12pt,a4wide,epsfig,amssymb]{article}

\title{
{\Large \bf 
Neutrino mass matrix solutions and neutrinoless double beta
decay\\}
\vspace*{0.8cm}
}

\author{T. Hambye
\\[0.5cm]
\normalsize  {\it Centre de Physique Th\'eorique,
CNRS Luminy,}\\[-0.15cm] \normalsize{\it Case 907, 13288 Marseille, France}
\\[2cm]
}

\date{}
\begin{document}
\maketitle
\thispagestyle{empty}
\vspace*{+0.2cm}
\begin{abstract}
We present a determination of the neutrino mass matrix
which holds for values of the neutrinoless double beta decay 
effective mass $m_{ee}$ larger than the
neutrino mass differences.
We find eight possible solutions and discuss for each one the corresponding
neutrino mass eigenvalues and zero texture. A minimal structure of 
the perturbations to add to these zero textures to recover the full mass matrix
is also determined. Implications for neutrino hot dark matter are
discussed for each solution.
\end{abstract}

\vspace*{\fill}
\noindent

\newpage
%
%
%
\noindent
{\bf 1)} \underline{\bf Introduction}: The neutrino mass 
matrix in the three neutrino framework 
has been severely constrained over the last few years
by atmospheric \cite{atm} and solar \cite{sol} neutrino experiments, and 
by recent positive indications from laboratory experiments
\cite{k2k,lsnd}. In the basis in which the charged lepton mass 
matrix is diagonal, the neutrino mass matrix contains all the
information about the mixing schemes in the leptonic sector. 
The mixing angle required for the atmospheric neutrinos appears to 
be maximal or very close to maximal. This is one of the main 
constraints on the neutrino mass matrix, when combined with the
mass squared difference required for atmospheric neutrinos
$\delta m^2_{atm}$,
which is in the range $(1.5-4.8) \cdot 10^{-3}$\cite{foglilisi}.
There are a few solutions to the solar neutrinos, which give
mass squared differences $\delta m^2_{sol}$ and mixing angle $\theta$ 
with $\nu_e$
which can lie in various ranges (see e.g. Ref.~\cite{bahcallnov}). 
The reactor experiment CHOOZ 
gives in addition a strong upper limit on the $U_{e3}$ element of
the neutrino mixing matrix \cite{chooz}. 
Besides these five constraints on the neutrino mass matrix, 
a sixth one would be given by 
the measurement of the effective
mass $m_{ee}$ in neutrinoless double beta decay.
This parameter would provide a scale to the neutrino mass matrix. 
Very recently a positive indication of neutrinoless double
has been reported
with $m_{ee}$ 
in the following range \cite{klap}:
\begin{equation}
m_{ee}= (0.05-0.86) \, \mbox{eV} 
\,\,\,(\mbox{at 95}\% \,\,\mbox{c.l.})
\label{meeexp}
\end{equation}
and with a best fit for $m_{ee}=0.39$~eV.
In the following 
we study what could be the consequences for the neutrino mass matrix 
of a value of $m_{ee}$ within 
the range of Eq.~(\ref{meeexp}). 
By considering this range we study 
more generally what would be the effect on the mass matrix
of an effective mass
$m_{ee}$ determined with a value 
greater than the neutrino mass differences and smaller than the
dark matter constraints (see
below). This range covers also the values of 
$m_{ee}$ which could be further tested experimentally in a relatively
near future \cite{experi}.
Since the scale $m_{ee}$ is then higher than the mass squared differences 
required by the neutrino oscillation results, this implies a
degenerate or partially degenerate solution \cite{klapsark}. The
hierarchical solution is not allowed by a value of $m_{ee}$ within the 
range of Eq.~(\ref{meeexp}). An 
inverse hierarchy solution
is still possible but only for $m_{ee}$ close to the lower edge of this
range.

Considering the fact that the atmospheric mixing is close to maximal,
several texture mass matrices have been constructed
\cite{altardelanazione}, including in particular the ones 
where the mass squared difference
$\delta m^2_{atm}$ is considered to be the dominant entry in the mass
matrix and where the solar neutrino solutions
could be added as perturbations. These textures are very
useful to construct models, where the zeroes could have their
origin in some symmetry of the model, so that the perturbations
come from the symmetry breaking. For a value of $m_{ee}$ within the range
of Eq.~(\ref{meeexp}),
$\delta m^2_{atm}$ would likely 
becomes a perturbation 
in the neutrino mass matrix with respect to the scale $m_{ee}$.
This would be an important information for the determination
of the texture of the mass matrix. 
Moreover, given a value of $m_{ee}$ in the range of Eq.~(\ref{meeexp}) 
the neutrino contribution to the
hot dark matter would be significant. With these assumptions we
now attempt to look into this problem. Instead of starting
the analysis from the possible textures, we start with the
experimental inputs and then try to determine the neutrino
mass matrices and in the process get the possible textures. 

In the following we shall work in the basis in which the 
charged lepton mass matrix is
diagonal. The physical states $| \nu_\alpha > $, ($\alpha = e, \mu,
\tau$), which enter in the charged current, are related to
the mass eigenstates $| \nu_i >$, (with masses $m_i$, $i = 1,2,3$),
by the mixing matrix
\begin{equation}
|\nu_\alpha > = U_{\alpha i} | \nu_i > 
\end{equation}
and the effective mass entering in the neutrinoless double beta 
decay is
\begin{equation}
m_{ee} =(M_\nu)_{ee}=(M_\nu)_{11}=\sum_i U_{ei}^2 m_i .
\end{equation}
$(M_\nu)_{11}$ is the $(11)$ element of the neutrino mass matrix
in the flavour basis. 

In this article we do not consider the imaginary part of
the mass matrix assuming that there is no CP violation, or very
small CP violation. 
In this limit, we 
can parametrize the mass matrix in the flavor basis 
as a function of the three real mass eigenvalues 
and the three angles coming from the mixing matrix $U_{\alpha i}$.
For simplicity and also as a first step of a more involved analysis we
consider in addition 
only the case with exact maximal mixing between the $\nu_\mu$
and $\nu_\tau$ (motivated by the best fit value from SuperKamiokande)
and we consider $U_{e3}=0$ (motivated by the result of the CHOOZ experiment).
In this framework we determine what 
are the effects 
on these 
six mass matrix parameters
of the five experimental constraints mentioned above taking
a given value of $m_{ee}$.
We find that there are eight different solutions for the mass matrix, each 
one being determined by these six constraints.
For each solution we display the corresponding zero texture and show how 
the perturbations arise around these zero textures to give the full
mass matrix. We also discuss the consequences for hot
dark matter for the different solutions. 
\vspace{2mm}

\noindent
{\bf 2)} \underline{\bf The neutrino mass matrix solutions}: 
In a general way the diagonal neutrino mass matrix can be written as:
\begin{equation}
M_\nu^{diag} = \left( \begin{array} {c@{\quad}c@{\quad}c@{\quad}c}
m_1 & 0 & 0  \\
0 & m_2 e^{i \phi_2} & 0 \\
0 & 0 &  m_3 e^{i \phi_3} 
\end{array} \right),
\end{equation}
here $\phi_{2,3}$ are the Majorana phases, which introduces CP
violation in lepton number violating processes. 
The mixing matrix $U_{\alpha i}$ may be parametrized as
\begin{equation}
U = \pmatrix{ c_1 c_3 & - s_1 c_3 & -s_3 e^{- i \phi} \cr
s_1 c_2 - c_1 s_2 s_3 e^{ i \phi} & c_1 c_2 + s_1 s_2 s_3 e^{ i \phi} &
- c_3 s_2 \cr s_1 s_2 + c_1 c_2 s_3 e^{ i \phi} &
c_1 s_2 - s_2 c_2 s_3 e^{ i \phi} & c_2 c_3 } ,
\end{equation}
where, $s_i = \sin \theta_i$ and $c_i = \cos \theta_i$.
This depends on 3 mixing angles $\theta_i$ and one 
CP-violating phase $\phi$. 
Thus there are 9 parameters in the general mass matrix 
in the flavor basis \cite{cabib}. 

The atmospheric neutrinos require an almost maximal mixing between
the flavour states $\nu_\mu$ and $\nu_\tau$. A very 
little mixing with the $\nu_e$ is not ruled out, but it is 
constrained by the CHOOZ result. Considering this constraint 
and the fact that the central value of the mixing angle for
atmospheric neutrinos is maximal, as
required by the SuperKamiokande result and supported by the
K2K preliminary data, we assume only maximal mixing 
$s_2 = 1/ \sqrt{2}$ between
the flavour states $\nu_\mu$ and $\nu_\tau$. Although the 
CHOOZ constraint allows a small $U_{e3}$, for construction of
a simple form of a mass matrix we assume it to be vanishing,
$s_3 = 0$. In this limit of $U_{e3}=0$  
the CP-violating effect from the phase $\phi$ in
$U$ vanishes and we can neglect its effect.
We can then write down the mixing matrix in terms of the
mixing angle $\theta = \theta_1$ required for the solar neutrinos solutions as
\begin{equation}
{U} = \left( \begin{array} {c@{\quad}c@{\quad}c@{\quad}c}
c & -s & 0  \\
s/\sqrt{2} & c/\sqrt{2}  & -1/\sqrt{2} \\
s/\sqrt{2} & c/\sqrt{2} & 1/\sqrt{2}  
\end{array} \right),
\end{equation}
with $c = \cos{\theta}$, $s = \sin{\theta}$.

This leads to the following
mass matrix in the $\nu_e$, $\nu_\mu$, $\nu_\tau$ flavor basis: 
\begin{equation}
M_\nu = \left( \begin{array} {c@{\quad}c@{\quad}c@{\quad}c}
c^2 m_1+s^2 m_2 & \frac{cs}{\sqrt{2}}(m_1-m_2) &  \frac{cs}{\sqrt{2}}(m_1-m_2)\\
\frac{cs}{\sqrt{2}}(m_1-m_2) & \frac{s^2 m_1 + c^2 m_2 +m_3}{2} & 
\frac{s^2 m_1+ c^2 m_2 -m_3}{2} \\
\frac{cs}{\sqrt{2}}(m_1-m_2) &  \frac{s^2 m_1 + c^2 m_2 -m_3}{2} & 
\frac{s^2 m_1 + c^2 m_2 +m_3}{2}
\end{array} \right),
\label{Mtotal}
\end{equation}

In the following we disregard the effect of the Majorana phases $\phi_{2,3}$ of 
the diagonal mass matrix. 
These CP violation
does not enter in any neutrino oscillation experiments. 
We will consider only real (positive or negative)
mass eigenvalues. By considering both positive and negative mass eigenvalues
we consider the extreme cases, and the case with non-zero phases are 
intermediate cases between these two extremes. The results with 
phases can be obtained as a simple 
generalization of what we present here.
Under these assumptions, to cover all physical distinct 
possibilities without loss of generality, 
it is sufficient to take $m_{ee}$ positive
and $|m_2| > |m_1|$ by definition and to consider the range
$0^\circ < \theta < 90^\circ$. Then from 
the value of $\delta m^2_{atm}$, $\delta m^2_{sol}$,
$\theta$, and $m_{ee}$ we can determine the mass matrix. There are 8 solutions.
To find them we have first to solve the equations:
\begin{equation}
m_2^2-m_1^2=\delta m^2_{sol}
\end{equation}
\begin{equation}
m_{ee}=m_1 c^2+ m_2 s^2
\end{equation}
This gives two sets of solutions: 
\begin{eqnarray}
m_1^{A}&=&m_{ee}-\frac{s^2}{2} \frac{\delta m^2_{sol}}{m_{ee}} \label{m1A}\\
m_2^{A}&=&m_{ee}+\frac{c^2}{2} \frac{\delta m^2_{sol}}{m_{ee}} \label{m2A}
\end{eqnarray}
and
\begin{eqnarray}
m_1^{B}&=&\frac{m_{ee}}{\cos{2 \theta}}+\frac{s^2}{2} 
\frac{\delta m^2_{sol}}{m_{ee}}\\
m_2^{B}&=&-\frac{m_{ee}}{\cos{2 \theta}}-\frac{c^2}{2} 
\frac{\delta m^2_{sol}}{m_{ee}}
\end{eqnarray}
where one can safely keep only the first order 
in $\delta m^2_{sol}/m_{ee}^2$ as we did here.
Then we can determine the mass $m_3$ from the equation:
\begin{equation}
m_3^2=\frac{m_1^2+m_2^2}{2} \pm \delta m^2_{atm}
\end{equation}
From this we find four solutions with the set $m^A_{1,2}$: 
\begin{eqnarray}
m_3^{A_1,A_2}&=& \sqrt{m_{ee}^2
\pm \delta m^2_{atm}} \label{m3A1}\\
m_3^{A_3,A_4}&=&- \sqrt{m_{ee}^2 
\pm \delta m^2_{atm}} \label{m3A3}
\end{eqnarray}
and four 
solutions with the set $m^B_{1,2}$:
\begin{eqnarray}
m_3^{B_1,B_2}&=& -\sqrt{\frac{m_{ee}^2}{\cos^2{2 \theta}} 
\pm \Delta m^2_{atm}} \label{m3B1}\\
m_3^{B_3,B_4}&=& +\sqrt{\frac{m_{ee}^2}{\cos^2{2 \theta}} 
\pm \Delta m^2_{atm}} \label{m3B3}
\end{eqnarray}
Here in $m_3$ we kept only the
correction in $\delta m^2_{atm}/m_{ee}^2$ and neglected 
the tiny correction from $\delta m^2_{sol}$.
\vspace{2mm}

\noindent
{\bf 3)} \underline{\bf Neutrino mass numerical values}: Putting 
Eqs.~(\ref{m1A})-(\ref{m3B3})
in Eq.~(\ref{Mtotal}), we can directly build the eight possible matrices
and substitute the values of $m_{ee}$, $\delta m^2_{atm}$, 
$\delta m^2_{sol}$ and $\theta$ to obtain the possible matrices. 
To illustrate the differences between the various solutions,
we give in Table 1 the values of the masses $m_{1,2,3}$ obtained in the
eight cases, 
taking, as a representative value of $m_{ee}$, the central 
value from \cite{klap},
$m_{ee}=0.39$~eV, and taking  
for the other parameters the following central values in the case 
of the LMA solution: 
$\delta m^2_{atm}=2.9 \cdot 10^{-3}$~eV$^2$ \cite{foglilisi},
$\delta m^2_{sol}=3.7 \cdot 10^{-5}$~eV$^2$ \cite{bahcallnov}, 
$\tan^2{\theta}=0.37$ \cite{bahcallnov}.
The uncertainties on these masses 
due to the experimental uncertainties on these 
four parameters are also given in Table 1. 
To get them we took the following ranges: $\delta m^2_{atm}=(1.5-4.8) 
\cdot 10^{-3}$~eV$^2$ \cite{foglilisi}, 
$\delta m^2_{sol}=(0.2-3) \cdot 10^{-4}$~eV$^2$ \cite{bahcallnov},
$\tan^2{\theta}=0.2-0.8$ \cite{bahcallnov}, $m_{ee}=(0.05-0.86)$~eV \cite{klap}.
The uncertainties on $U_{e3}$ and the atmospheric angle also 
induce an error, which, however, is moderate and which we didn't consider.
We will discuss these central values and ranges in sections 5 and 6.
\begin{table}[t]
\begin{eqnarray*}
\begin{array}{|c|c||c|c|}
\hline
\rule{0cm}{5mm}
& \,\,m_1=0.390 \,\, &  & m_1=0.848
\\[0.5mm]
& \,\,m_2=0.390 \,\, &  &m_2 =-0.848
\\[0.5mm]
A_1&\,\,m_3=0.394 \,\, & B_1 & m_3=-0.850
\\[0.5mm]
& \,\,0.050\leq m_1\leq 0.86 \,\, &  & 0.0750\leq m_1\leq 7.74
\\[0.5mm]
& \,\,0.050\leq m_2 \leq 0.86\,\, &  & -7.74 \leq m_2 \leq -0.0750
\\[0.5mm]
&\,\,0.0632\leq m_3 \leq 0.8628\,\, &  &-7.74\leq m_3 \leq -0.0845
\\[0.5mm]
\hline \hline
& \,\,m_1=0.390 \,\, &  & m_1=0.848
\\[0.5mm]
& \,\,m_2=0.390 \,\, &  &m_2 =-0.848
\\[0.5mm]
A_2&\,\,m_3=0.386 \,\, & B_2 & m_3=-0.846
\\[0.5mm]
& \,\, 0.049\leq m_1\leq 0.860\,\, &  & 0.0750\leq m_1\leq 7.74
\\[0.5mm]
& \,\, 0.050\leq m_2 \leq 0.860 \,\, &  & -7.74 \leq m_2\leq -0.075
\\[0.5mm]
&\,\,0\leq m_3 \leq 0.859\,\, &  &-7.74\leq m_3 \leq-0.029 
\\[0.5mm]
\hline 
\end{array}
\end{eqnarray*}
\caption{Central values and allowed ranges of the mass eigenvalues $m_{1,2,3}$
(in eV).
For the solutions $A_3$, $A_4$, $B_3$, $B_4$, the mass eigenvalues are the same
as that of the solutions $A_1$, $A_2$, $A_3$, $A_4$, respectively,
except for the sign
of $m_3$ which has to be reversed. We put several digits just to show the effect
of $\delta m^2_{atm}$ which is different for the various solutions.
}
\end{table}

For the SMA solution, although this solution is now almost ruled 
out \cite{bahcallnov}, the same calculation can be done and for completeness 
we give in Table~2 the values of $m_{1,2,3}$ we obtain considering 
the following central values from Ref.~\cite{bahcallnov}:
$\delta m^2_{sol}=5.2 \cdot 10^{-6}$~eV$^2$ and 
$\tan^2{\theta}=1.8 \cdot 10^{-3}$. We don't give here any range because 
the central values for the SMA solution have been obtained 
in Ref.~\cite{bahcallnov} beyond the 3-$\sigma$ level.
The results for the LOW and VAC solar neutrino solutions 
can be obtained in the same way
but we don't present it because
they are in general difficult to accommodate with
a value of $m_{ee}$ larger than the neutrino mass 
differences \cite{klapsark}. See 
however counter-examples in Ref.~\cite{strumia}.
\begin{table}[t]
\begin{eqnarray*}
\begin{array}{|c|c||c|c|}
\hline
\rule{0cm}{5mm}
& \,\,m_1=0.390 \,\, &  & m_1=0.391
\\[0.5mm]
A_1& \,\,m_2=0.390 \,\, & B_1 &m_2 =-0.391
\\[0.5mm]
&\,\,m_3=0.394 \,\, &  & m_3=-0.395
\\[0.5mm]
\hline \hline
& \,\,m_1=0.390 \,\, &  & m_1=0.391
\\[0.5mm]
A_2 & \,\,m_2=0.390 \,\, & B_2 & m_2 =-0.391
\\[0.5mm]
&\,\,m_3=0.386 \,\, & & m_3=-0.388
\\[0.5mm]
\hline 
\end{array}
\end{eqnarray*}
\caption{Central values of the mass eigenvalues $m_{1,2,3}$
(in eV) for the SMA solution.
For the solutions $A_3$, $A_4$, $B_3$, $B_4$, the mass eigenvalues are the same
than for the solutions $A_1$, $A_2$, $A_3$, $A_4$ respectively 
except for the sign
of $m_3$ which has to be reversed.
}
\end{table}
\vspace{2mm}

\noindent
{\bf 4)} \underline{\bf Zero textures and perturbations}: To 
understand better the structure of these matrices and also for the 
purpose of model building it is interesting to write these matrices in 
the form of textures zero plus the perturbations. 
This will also allow us
to compare our results with the textures predicted from different
considerations \cite{altardelanazione}. In the following we shall consider 
the degenerate solution assuming that $m_{ee}$ is larger than the 
neutrino mass differences.
For the inverse hierarchy solution 
we will just give a few comments
in sections 5 and 6.

In the case 
where the mass eigenvalues are almost degenerate with the mass squared 
differences to be small compared to the eigenvalues, i.e.
$|m_{1,2,3}| \gg \delta m^2_{atm} \gg \delta m^2_{sol}$, the 
parameters $\delta m^2_{atm}$ and $\delta m^2_{sol}$ can be considered 
as perturbations with respect to $m_{ee} \sim |m_{1,2,3}| $.
In this case we get the following zero textures and
perturbations for the different solutions:
\begin{eqnarray}
M_{A_1,A_2} &=&
m_{ee} \left( \begin{array} {c@{\quad}c@{\quad}c@{\quad}c}
1 & 0 & 0  \\
0 & 1 &
 0  \\ 
0 &  0 &  
1
\end{array} \right) \pm
m_{ee} \frac{\varepsilon_{atm}}{4} 
\left( \begin{array} {c@{\quad}c@{\quad}c@{\quad}c}
0 & 0 & 0  \\
0 & 1 &
 -1 \\ 
0 &  -1 &  
1
\end{array} \right)\nonumber\\&+&
m_{ee} \frac{\varepsilon_{sol}}{8} 
\left( \begin{array} {c@{\quad}c@{\quad}c@{\quad}c}
0 & -{\sqrt{2}}{\sin{2 \theta}} & -{\sqrt{2}}{\sin{2 \theta}} \\
-{\sqrt{2}}{\sin{2 \theta}}  & {2 \cos{2 \theta}}
&  {2 \cos 2 \theta} \\
-{\sqrt{2}}{\sin{2 \theta}} &  
{2 \cos 2 \theta} & 
{2 \cos{2 \theta}} 
\end{array} \right), \label{MA1}
\end{eqnarray}

\begin{eqnarray}
M_{A_3,A_4} &=& m_{ee} 
\left( \begin{array} {c@{\quad}c@{\quad}c@{\quad}c}
1 & 0 & 0  \\
0 & 0 & 1 \\ 
0 & 1 & 0
\end{array} \right)
\pm m_{ee}\frac{\varepsilon_{atm}}{4} 
\left( \begin{array} {c@{\quad}c@{\quad}c@{\quad}c}
0 & 0 & 0  \\
0 & -1& +1  \\ 
0 & +1&  -1
\end{array} \right)\nonumber\\
&+& m_{ee} \frac{\varepsilon_{sol}}{8}
\left( \begin{array} {c@{\quad}c@{\quad}c@{\quad}c}
0 &  
-{\sqrt{2}}{\sin{2 \theta}} & 
-{\sqrt{2}}{\sin{2 \theta}} \\
-{\sqrt{2}}{\sin{2 \theta}} & 
{2 \cos{2 \theta}} 
& 
{2 \cos 2 \theta}\\
-{\sqrt{2}}{\sin{2 \theta}} &  
{2 \cos 2 \theta}& 
{2 \cos{2 \theta}}
\end{array} \right), \label{MA3}
\end{eqnarray}

\begin{eqnarray}
M_{B_1,B_2} &=& \frac{m_{ee}}{ \cos 2 \theta} 
\left( \begin{array} {c@{\quad}c@{\quad}c@{\quad}c}
 \cos 2 \theta &
\frac{\sin{2 \theta}}{\sqrt{2}} &  
\frac{\sin{2 \theta}}{\sqrt{2}} \\  
\frac{\sin{2 \theta}}{\sqrt{2}} &  
-\frac{1}{2} -\frac{\cos{2 \theta}}{2} 
&  \frac{1}{2} - \frac{\cos 2 \theta}{2} \\
\frac{\sin{2 \theta}}{\sqrt{2}} & \frac{1}{2} - \frac{\cos 2 \theta}{2} & 
-\frac{1}{2} - \frac{\cos 2 \theta}{2} 
\end{array} \right) \nonumber \\ &\pm& 
m_{ee} \cos{2 \theta} \frac{\varepsilon_{atm}}{4}
\left( \begin{array} {c@{\quad}c@{\quad}c@{\quad}c}
0 & 0 & 0  \\
0 &  - 1 &  +1\\
0 & +1
& -1
\end{array} \right)
 \nonumber \\&+&
m_{ee} \frac{\varepsilon_{sol}}{8}
\left( \begin{array} {c@{\quad}c@{\quad}c@{\quad}c}
0 & 
\sqrt{2} {\sin{2 \theta}} & \sqrt{2} {\sin{2 \theta}} \\
\sqrt{2} {\sin{2 \theta}} & -2 \cos 2 \theta &
 -2 \cos 2 \theta \\ 
\sqrt{2} {\sin{2 \theta}} &  -2 \cos 2 \theta &
 -2 \cos 2 \theta 
\end{array} \right), \label{MB1}
\end{eqnarray}

\begin{eqnarray}
M_{B_3,B_4} &=& \frac{m_{ee}}{ \cos 2 \theta} 
\left( \begin{array} {c@{\quad}c@{\quad}c@{\quad}c}
 \cos 2 \theta &
\frac{\sin{2 \theta}}{\sqrt{2}} &  
\frac{\sin{2 \theta}}{\sqrt{2}} \\  
\frac{\sin{2 \theta}}{\sqrt{2}} &  
\frac{1}{2} -\frac{\cos{2 \theta}}{2} 
&  -\frac{1}{2} - \frac{\cos 2 \theta}{2} \\
\frac{\sin{2 \theta}}{\sqrt{2}} & -\frac{1}{2} - \frac{\cos 2 \theta}{2} & 
\frac{1}{2} - \frac{\cos 2 \theta}{2} 
\end{array} \right) \nonumber \\ &\pm& 
m_{ee} \cos{2 \theta} \frac{\varepsilon_{atm}}{4} 
\left( \begin{array} {c@{\quad}c@{\quad}c@{\quad}c}
0 & 0 & 0  \\
0 &   1 & -1\\
0 & -1 &1
\end{array} \right) \nonumber \\
&+& m_{ee} \frac{\varepsilon_{sol}}{8}
\left( \begin{array} {c@{\quad}c@{\quad}c@{\quad}c}
0& \sqrt{2} \sin{2 \theta} & \sqrt{2} \sin{2 \theta} \\ 
\sqrt{2} \sin{2 \theta} &  - 2 \cos{2 \theta} & 
 - 2 \cos{2 \theta} \\  \sqrt{2} \sin{2 \theta} &
 - 2 \cos{2 \theta}&  - 2 \cos{2 \theta}
\end{array} \right),  \label{MB3}
\end{eqnarray}
Here $\varepsilon_{sol} \equiv \delta m^2_{sol}/m^2_{ee}$ and
$\varepsilon_{atm} \equiv \delta m^2_{atm}/m_{ee}^2$. 
The corresponding mass eigenvalues are given in Table 3. Note that the various
solutions differ from each other by their zero texture
and by prefactors in the perturbations. Except for these different
prefactors the various perturbations have the same matrix structure
for all the mass matrix solutions.
\begin{table}[t]
\begin{center}
\begin{tabular}{|c||c|c|}
\hline
\rule{0mm}{5mm}
 & $M^\nu_{(0)}$& $m_{1,2,3}$\\[1mm] \hline \hline &&\\[-3mm]
$A_{1,2}$ &  
$
{\footnotesize  m_{ee} \, \left( \begin{array} {c@{\quad}c@{\quad}c@{\quad}c}
1 & 0 & 0  \\
0 &  1 &  0\\
0 & 0
& 1
\end{array} \right)}
$
&
{\footnotesize $\begin{array}{c} 
m_1=m_{ee}(1-s^2\frac{\varepsilon_{sol}}{2})\\
m_2=m_{ee}(1+c^2\frac{\varepsilon_{sol}}{2})\\
m_3=m_{ee}(1\pm
\frac{\varepsilon_{atm}}{2}) \end{array}$}
\\[5mm] &&\\[-3mm]
\hline &&\\[-3mm]
$A_{3,4}$ &  
$
{\footnotesize  m_{ee} \, \left( \begin{array} {c@{\quad}c@{\quad}c@{\quad}c}
1 & 0 & 0  \\
0 &  0 &  1\\
0 & 1
& 0 
\end{array} \right)}
$
&
{\footnotesize $\begin{array}{c} 
m_1=m_{ee}(1-s^2\frac{\varepsilon_{sol}}{2})\\
m_2=m_{ee}(1+c^2\frac{\varepsilon_{sol}}{2})\\
m_3=-m_{ee}(1\pm
\frac{\varepsilon_{atm}}{2}) \end{array}$ }
\\[5mm] &&\\[-3mm]
\hline &&\\[-3mm]
$B_{1,2}$ &  
{\footnotesize $
\tilde{m} \, \left( \begin{array} {c@{\quad}c@{\quad}c@{\quad}c}
 \cos 2 \theta &
\frac{\sin{2 \theta}}{\sqrt{2}} &  
\frac{\sin{2 \theta}}{\sqrt{2}} \\  
\frac{\sin{2 \theta}}{\sqrt{2}} &  
-\frac{1}{2} -\frac{\cos{2 \theta}}{2} 
&  \frac{1}{2} - \frac{\cos 2 \theta}{2} \\
\frac{\sin{2 \theta}}{\sqrt{2}} & \frac{1}{2} - \frac{\cos 2 \theta}{2} & 
-\frac{1}{2} - \frac{\cos 2 \theta}{2} 
\end{array} \right)
$ }
&
{\footnotesize $\begin{array}{c} 
m_1=\tilde{m}(1+s^2 \cos{2\theta}
\frac{\varepsilon_{sol}}{2})\\
m_2=-\tilde{m}(1+c^2 \cos{2 \theta}
\frac{\varepsilon_{sol}}{2})\\
m_3=-\tilde{m}(1\pm
\cos^2{2 \theta} \frac{\varepsilon_{atm}}{2}) \end{array}$}
\\[5mm] &&\\[-3mm]
\hline &&\\[-3mm]
$B_{3,4}$ &  
{\footnotesize $
\tilde{m} \, \left( \begin{array} {c@{\quad}c@{\quad}c@{\quad}c}
 \cos 2 \theta &
\frac{\sin{2 \theta}}{\sqrt{2}} &  
\frac{\sin{2 \theta}}{\sqrt{2}} \\  
\frac{\sin{2 \theta}}{\sqrt{2}} &  
\frac{1}{2} -\frac{\cos{2 \theta}}{2} 
&  -\frac{1}{2} - \frac{\cos 2 \theta}{2} \\
\frac{\sin{2 \theta}}{\sqrt{2}} & -\frac{1}{2} - \frac{\cos 2 \theta}{2} & 
\frac{1}{2} - \frac{\cos 2 \theta}{2} 
\end{array} \right) 
$}
&
{\footnotesize $\begin{array}{c} 
m_1=\tilde{m}(1+s^2 \cos{2\theta}
\frac{\varepsilon_{sol}}{2})\\
m_2=-\tilde{m}(1+c^2 \cos{2 \theta}
\frac{\varepsilon_{sol}}{2})\\
m_3=\tilde{m}(1\pm
\cos^2{2 \theta} \frac{\varepsilon_{atm}}{2}) \end{array}$}
\\[8mm]
\hline 
\end{tabular}
\caption{Zero textures and neutrino mass eigenvalues 
at first order in $\varepsilon_{atm}$ and $\varepsilon_{sol}$ 
corresponding to the solutions $A_{1,2,3,4}$ and 
to $B_{1,2,3,4}$.
}
\end{center}
\end{table}

\vspace{2mm}

\noindent
{\bf 5)} \underline{\bf The $A_{1,2,3,4}$ solutions}: First let 
us discuss the solutions $A_1$ to $A_4$:
\begin{itemize}
\item \underline{{\it Mass eigenvalues}}: 
For these solutions the absolute values of the masses of 
the 3 eigenstates 
coincide with $m_{ee}$ modulo small 
variations (proportional to $\delta m^2_{atm}$ and
$\delta m^2_{sol}$). This explains why in Table 1 the upper limit of the 
neutrino 
mass ranges reproduce to a good approximation the upper limit on $m_{ee}$.
For low values of $m_{ee}$ (around 0.05~eV in Eq.~(\ref{meeexp})), 
the neutrino masses 
$m_{1,2}$ are still given to a good approximation 
by $m_{ee}$ but the $|m_3|$ value can differ 
substantially from the 
one of $m_{ee}$. This is due to the fact that in this case
$\delta m^2_{atm}$ is not 
small anymore with respect to $m_{ee}$ and $|m_3|$ 
could be as low as 0 eV for 
the solutions $A_{2,4}$ (where $\delta m^2_{atm}$ 
and $m_{ee}$ come with opposite signs in $m_3$). This later case corresponds 
to the inverse hierarchy scenario.

\item \underline{{\it Dark matter}}: For a value of $m_{ee}$ such as
the central 
value $m_{ee}=0.39$~eV \cite{klap} or above, the
solutions $A_{1,2,3,4}$ could be relevant for dark matter.
$|m_1|=|m_2|=|m_3|=m_{ee}=0.39$~eV (or the upper value 0.86 eV 
in Eq.~(\ref{meeexp})) 
corresponds, in units of the critical density, to a value of 
$\Omega_\nu h^2 \sim \sum |m_i|/(91.5 \mbox{eV})$ 
around $\sim 1.5 \%$ ($\sim 3 \%$) with $h$ the dimensionless 
Hubble constant. Note that the 
value $\Omega_\nu h^2 \sim 3 \%$ turns out to be about the upper bound
allowed by fits from observed galaxy clustering and the Lyman $\alpha$ 
forest (see e.g. \cite{barger}-\cite{sdss} and the 
discussion in \cite{klapsark}).

\item \underline{{\it Textures and perturbations in the degenerate case}}: 
For the cases $A_{1,2}$ and $A_{3,4}$ we recognize 
the textures C0 and C3
of Altarelli-Feruglio, respectively \cite{altardelanazione}(naturally 
leading to both simple maximal 
mixing or double maximal mixing). 
From Eqs.~(\ref{Mtotal})-(\ref{m3B3}) one can observe also  
how the perturbations arise with respect to these zero textures
to give the full mass matrix. 
Note that the structure of the perturbations
due to $\delta m^2_{atm}$ is quite simple, which is interesting 
for model building. Note also that in case $A_{1,2}$ it is 
this perturbation which is responsible for the maximal atmospheric mixing.
In case $A_{3,4}$ the maximal mixing is already present 
for $\delta m^2_{atm}=0$. The structure of the perturbation 
from $\delta m^2_{sol}$ is more complicated but remains rather
simple especially if $\theta$ takes a 
specific value (i.e. for example for $\theta=0$ or $\theta=\pi/4$).
In all cases $A_1$ to $A_4$ the solar mixing is due to the perturbation
proportional to $\delta m^2_{sol}$. 
The solar angle even if maximal 
is a perturbation because all the terms depending on it are multiplied by 
$\delta m^2_{sol}$. 
There is no solar mixing 
if $\delta m^2_{sol}$ vanishes.

\item \underline{{\it The inverse hierarchy case}}: 
If $m_{ee}$ turns out to be not much larger than the 
lower edge of the experimental
range in Eq.~(\ref{meeexp}), i.e. $m_{ee}=0.05$~eV, we could have an 
inverse hierarchy where $|m_3| << |m_{1,2}|$ which means,
since $m_{ee}^2 >> \delta m^2_{sol}$, that $|m_3| << |m_{ee}|$.
Such a situation requires a cancellation in $m_3$, Eqs.~(\ref{m3A1}) and 
(\ref{m3A3}), of 
the $m_{ee}$ term with
the term proportional to $\delta m^2_{atm}$. This is 
possible only with solutions $A_{2,4}$ 
where $\delta m^2_{atm}$ comes with an opposite sign with respect to the term
in $m_{ee}^2$ in $m_3$. For these solutions this requires
$m^2_{ee} \sim \delta m^2_{atm}$. 
For the central values
$\delta m^2_{atm}=2.9 \cdot 10^{-3}$~eV$^2$ \cite{foglilisi} and 
$\tan^2{\theta}=0.37$ \cite{bahcallnov} this requires 
$m_{ee}\sim 0.054$~eV.
Varying 
$\delta m^2_{atm}$ within 
the range $(1.5-4.8) \cdot 10^{-3}$~eV$^2$ this requires: 
\begin{equation}
4 \cdot 10^{-2} \,\, \mbox{eV} \, < m_{ee} < \, 7 \cdot 
10^{-2} \,\,\mbox{eV}\,\,\, \mbox{(for 
solutions $A_{2,4}$)} \label{meerange1} 
\end{equation}
This range is independent of the solar neutrino 
solution. Note also that we could develop in this case the mass matrix
considering $m_3$ and $\sqrt{\delta m^2_{sol}}$ as perturbations with
respect to $m_{ee}$ and $\sqrt{\delta m^2_{atm}}$. In this 
case the zero texture we obtain is 
\begin{equation}
\left( \begin{array} {c@{\quad}c@{\quad}c@{\quad}c}
1 & 0 & 0  \\
0 & 1/2 & 1/2 \\
0 & 1/2 & 1/2  
\end{array} \right),  
\end{equation}
that is to say the texture B2 of 
Ref.~\cite{altardelanazione} which can lead to both 
single and double maximal mixing.

\end{itemize}
\vspace{2mm}

\noindent
{\bf 6)} \underline{\bf The $B_{1,2,3,4}$ solutions}: Solutions $B_1$ to
$B_4$ are very different from $A_1$ to $A_4$ and more complicated 
because $m_1$ and $m_2$ here have opposite sign (c.f. Table 3):
\begin{itemize}
\item \underline{{\it Neutrino mass eigenvalues}}: Here 
in the limit $\delta m^2_{atm}=\delta m^2_{sol}=0$ 
the absolute values of the neutrino masses are not $m_{ee}$ (as with 
solutions $A_1$ to $A_4$) 
but turn out to be in fact $\tilde{m} \equiv m_{ee}/\cos{2 \theta}$. This 
explains the large
value of the neutrino masses obtained in Table 1 with these solutions.
Taking the central value
$\tan^2{ \theta} \sim 0.39$ \cite{bahcallnov}
for the LMA solution we then get for 
the neutrino masses $\tilde{m} \sim 2.2 \, m_{ee}$. For a value
of $m_{ee}=0.39$~eV this gives $\tilde{m} \sim 0.85$ eV.
Taking the upper limit $\tan^2{\theta} \sim 0.8$ 
\cite{bahcallnov} we 
get $\tilde m \sim 9 \, m_{ee}$ which, for $m_{ee}=0.86$~eV in Eq.~(\ref{meeexp}),
could result in a 
neutrino mass as large as 8 eV. However, a neutrino mass greater than
2.8 eV is excluded by the tritium beta decay experiment \cite{tritium}.

\item \underline{{\it Dark matter}}: These 
values $\tilde{m}=0.85$~eV and $\tilde{m}=8$~eV
correspond to $\Omega_\nu h^2 \sim 3 \%$.
and 
to $\Omega_\nu h^2 \sim 25 \%$ respectively.  
From Table~1 we then get the following range for solutions $B_{1,2,3,4}$:
\begin{equation}
0.001<\Omega_\nu h^2 < 25 \%
\end{equation}
Note that, as already mentioned above, 
a value of $\Omega_\nu h^2$ larger than $\sim3 \%$ 
which corresponds to $\sum m_i \sim 3$ eV is 
disfavoured by the results of galaxy clustering and Lyman 
$\alpha$ forest experiments \cite{barger}-\cite{sdss}. In case of the solutions
$B_{1,2,3,4}$, to have $\Omega_\nu h^2$ 
smaller than $\sim 3\%$ we need 
that the following 
constraint is fulfilled:
\begin{equation}
\tilde{m}=\frac{m_{ee}}{\cos{2 \theta}}< 1 \, \mbox{eV}
\end{equation}
For example this condition for $m_{ee}=0.86$~eV in Eq.~(\ref{meeexp}) 
requires $\tan^2{\theta} < 0.08$, 
which can be compared with the lower limit of the LMA solution which is
$\tan^2{\theta} \sim 0.2$ \cite{bahcallnov}. This means that a value 
of $m_{ee}$ as large as 0.86 eV with the solutions $B_{1,2,3,4}$
is disfavoured by dark matter fits. To obtain $\tilde{m} < 1$~eV with 
for example $\tan^2{2 \theta}=0.2, \, 0.39, \, 0.8$ we need
$m_{ee}<0.7$~eV, 0.4~eV, 0.1~eV respectively. Would we allow
$\Omega_\nu h^2$ to be as large as $10 \%$, these bounds would be 
multiplied obviously by a factor 3.
Note that for a $\Lambda$CHDM model with both hot and cold 
dark matter as well as a cosmological constant a value of 
$\tilde m < 0.5$ eV is favoured \cite{gasilk}. 
This already favours a lower
value for the neutrinoless double beta decay in this scheme. 
Furthermore, the CMB studies with MAP and PLANCK and the
Sloan Digital Sky Survey experiment (see 
e.g. \cite{eisenstein}-\cite{lopez}) will 
be able
to probe the mass range $\tilde m \sim 0.25 - 0.5$ eV. If the 
contribution to the hot dark matter from neutrinos is constrained
to be less than $\tilde m < 0.25$ eV, then this will mean that
this class of solutions will be further restricted and a lower value of
the neutrinoless double beta decay will be favoured. 
Note also that the fact that the LMA solution in the degenerate case
could be relevant for dark matter is well known \cite{vissani}. Here we  
emphasize that, taking all the experimental constraints
on $U_{e3}$, $\delta m^2_{atm}$, $\delta m^2_{solar}$ and taking
$m_{ee}$ within the range of Eq.~(\ref{meeexp}), all the solutions for 
the mass matrix, and especially the $B_{1,2,3,4}$ solutions lead 
naturally to such a rather important contribution 
to dark matter. Inversely the dark matter constraint prefers
a lower value of the $m_{ee}$ or a higher $\cos{2 \theta}$
for the solutions $B_{1,2,3,4}$.
Note also that for the SMA solution with the central 
values $\delta m^2_{sol}=5.2 \cdot 10^{-5}$~eV$^2$ and 
$\tan^2{\theta}=1.8 \cdot 10^{-3}$ we 
obtain $\tilde{m}=m_{ee}$ to a good approximation which 
means for $m_{ee}=0.39$~eV that $\Omega_\nu h^2= 1.5 \%$.

\item \underline{{\it Textures and perturbations in the degenerate case}}:
In contrast with solutions $A_{1,2,3,4}$ the 
dependence in $\theta$ is 
already non-vanishing even without any perturbation proportional to 
$\delta m^2_{sol}$ and $\delta m^2_{atm}$. So even in this 
limit there is in fact a "continuum" of zero textures depending on the 
value of $\theta$, see Eqs.~(\ref{MB1}) and (\ref{MB3}). 
This allows for a continuum of basis where the zero texture is diagonal,
i.e. of basis which can be considered as natural in the symmetry limit.
Eventually for some particular values of the 
angle $\theta$ the corresponding diagonal mass matrix will display
interesting properties. These considerations are beyond the scope 
of this article.

The 
structure of the perturbations appear to
be also more complicated than for $A_{1,2,3,4}$. In particular, now there is 
a dependence of the perturbation from $\delta m^2_{atm}$ in $\theta$, 
which would lead to the interpretation that this perturbation cannot 
be independent of the other perturbations, unlike in the solutions $A_{1,2,3,4}$
where they are independent.
In the continuum of 
possible zero textures for the $B_1$ to $B_4$ 
solutions, a natural choice is to take the textures with $\theta=0$ 
or $\theta= \pi/4$,  
and make a perturbation around these cases (for the SMA and 
the LMA respectively). In this case the deviation
of $\theta$ from these values can be considered as a perturbation. 
For $\theta=0$ 
we obtain: 
\begin{eqnarray}
M_{B_1,B_2} &=& m_{ee} \left( \begin{array} {c@{\quad}c@{\quad}c@{\quad}c}
1 & 0 &  0 \\  
0 &  -1 &  0\\
0 & 0 & -1
\end{array} \right) \nonumber \\ &+&
{m_{ee}}
\left( \begin{array} {c@{\quad}c@{\quad}c@{\quad}c}
0 & 
\frac{\tan{2 \theta}}{\sqrt{2}} &
\frac{\tan{2 \theta}}{\sqrt{2}} \\
\frac{\tan{2 \theta}}{\sqrt{2}} 
&  
\frac{1}{2}-\frac{1}{2 \cos{2 \theta}}
&  
-\frac{1}{2}+\frac{1}{2 \cos{2 \theta}}\\
\frac{\tan{2 \theta}}{\sqrt{2}}
& 
-\frac{1}{2}+\frac{1}{2 \cos{2 \theta}}
& 
\frac{1}{2}-\frac{1}{2 \cos{2 \theta}}
\end{array} \right)
\nonumber\\
&\pm& m_{ee} \cos{2 \theta}
\frac{\varepsilon_{atm}}{4}\left( \begin{array} {c@{\quad}c@{\quad}c@{\quad}c}
0 & 0 & 0  \\
0 &  - 1 &  +1\\
0 & +1
& -1
\end{array} \right) \nonumber \\ &+&
m_{ee} \frac{\varepsilon_{sol}}{8}
\left( \begin{array} {c@{\quad}c@{\quad}c@{\quad}c}
0 & \sqrt{2} \sin{2 \theta} & \sqrt{2} \sin{2 \theta} \\
\sqrt{2} \sin{2 \theta} & -2 \cos{2 \theta} 
& -2 \cos{2 \theta} \\ \sqrt{2} \sin{2 \theta} &
-2 \cos{2 \theta}  & -2 \cos{2 \theta} 
\end{array} \right),
\end{eqnarray}

\begin{eqnarray}
M_{B_3,B_4} &=& m_{ee} \left( \begin{array} {c@{\quad}c@{\quad}c@{\quad}c}
1 & 0 &  0 \\  
0 &  0 &  -1\\
0 & -1 & 0 
\end{array} \right) \nonumber \\ &+&
{m_{ee}} 
\left( \begin{array} {c@{\quad}c@{\quad}c@{\quad}c}
0 & 
\frac{\tan{2 \theta}}{\sqrt{2}} &
\frac{\tan{2 \theta}}{\sqrt{2}} \\
\frac{\tan{2 \theta}}{\sqrt{2}} 
&  
-\frac{1}{2}+\frac{1}{2 \cos{2 \theta}}
&  
\frac{1}{2}-\frac{1}{2 \cos{2 \theta}}\\
\frac{\tan{2 \theta}}{\sqrt{2}}
& 
\frac{1}{2}-\frac{1}{2 \cos{2 \theta}}
& 
-\frac{1}{2}+\frac{1}{2 \cos{2 \theta}}
\end{array} \right)
\nonumber \\&\pm&
m_{ee} \frac{\varepsilon_{atm}}{4} \cos{2 \theta}
\left( \begin{array} {c@{\quad}c@{\quad}c@{\quad}c}
0 & 0 & 0  \\
0 &   1 & -1\\
0 & -1 &1
\end{array} \right) \nonumber \\ &+& m_{ee} \frac{\varepsilon_{sol}}{8}
\left( \begin{array} {c@{\quad}c@{\quad}c@{\quad}c}
0 & \sqrt{2} \sin{2 \theta} & \sqrt{2} \sin{2 \theta} \\
\sqrt{2} \sin{2 \theta} & - 2 \cos{2 \theta} &- 2 \cos{2 \theta}\\
\sqrt{2} \sin{2 \theta} & - 2 \cos{2 \theta} &- 2 \cos{2 \theta}
\end{array} \right),
\end{eqnarray}
Here we recover the textures C1 and C2 of 
Altarelli-Feruglio \cite{altardelanazione}(for the 
single maximal mixing case) and the 
absolute neutrino mass in the limit $\delta m^2_{sol}=\delta m^2_{atm}=0$
is $m_{ee}/ \cos{2 \theta} \sim m_{ee}$.
The SMA solution could be obtained from this texture 0 but in this case,
as already said above,
the structure of the perturbation appears to be quite complicated, more 
complicated than from the texture in solutions $A_{1,2,3,4}$. 

We then consider the solution around $\theta=\pi/4$. In this case
there are interesting zero textures. As explained above, 
the absolute neutrino mass, in the limit $\delta m^2_{sol}= 
\delta m^2_{atm}=0$, is now given by $\tilde{m} \equiv m_{ee}/\cos{2 \theta}$. 
In this case $m_{ee}$ can be considered
as a perturbation with respect to $m$:
\begin{eqnarray}
{\cal M}_{B_1,B_2} &=& {\tilde{m} }
 \left( \begin{array} {c@{\quad}c@{\quad}c@{\quad}c}
0 & \frac{1}{\sqrt{2}} &  \frac{1}{\sqrt{2}} \\  \frac{1}{\sqrt{2}} &  
- \frac{1}{2} & \frac{1}{2} \\ \frac{1}{\sqrt{2}} & \frac{1}{2} & -\frac{1}{2} 
\end{array} \right) \nonumber \\ &+& {\tilde{m}}
 \left( \begin{array} {c@{\quad}c@{\quad}c@{\quad}c}
 \cos{2 \theta} &  \frac{ \sin{2 \theta} - 1}{\sqrt{2}}
& \frac{\sin{2 \theta} - 1}{\sqrt{2}} \\
 \frac{\sin{2 \theta} - 1}{\sqrt{2}} &
- \frac{\cos{2 \theta}}{2} &  - \frac{\cos{2 \theta}}{2} \\
\frac{ \sin{2 \theta} - 1}{\sqrt{2}} &
- \frac{\cos{2 \theta}}{2} &  -\frac{ \cos{2 \theta}}{2} 
\end{array} \right)\nonumber \\
&\pm& \tilde{m} \cos^2{2 \theta}
\frac{\varepsilon_{atm}}{4}\left( \begin{array} {c@{\quad}c@{\quad}c@{\quad}c}
0 & 0 & 0  \\
0 &  - 1 &  +1\\
0 & +1
& -1
\end{array} \right) \nonumber \\ &+&
\tilde{m} \cos{2 \theta} \frac{\varepsilon_{sol}}{8}
\left( \begin{array} {c@{\quad}c@{\quad}c@{\quad}c}
0 & \sqrt{2} \sin{2 \theta} & \sqrt{2} \sin{2 \theta} \\
\sqrt{2} \sin{2 \theta} & - 2 \cos{2 \theta} 
& -2 \cos{2 \theta} \\ \sqrt{2} \sin{2 \theta} &
-2 \cos{2 \theta}  & -2 \cos{2 \theta} 
\end{array} \right),
\label{MB1thetapi}
\end{eqnarray}

\begin{eqnarray}
{\cal M}_{B_3,B_4} &=& {\tilde{m} }
 \left( \begin{array} {c@{\quad}c@{\quad}c@{\quad}c}
0 & \frac{1}{\sqrt{2}} & \frac{1}{\sqrt{2}} \\ \frac{1}{\sqrt{2}} & \frac{1}{2} & -\frac{1}{2} \\
\frac{1}{\sqrt{2}} & -\frac{1}{2} & \frac{1}{2} 
\end{array} \right) \nonumber \\ &+& {\tilde{m} }
 \left( \begin{array} {c@{\quad}c@{\quad}c@{\quad}c}
\cos{2 \theta} & \frac{\sin{2 \theta} - 1}{\sqrt{2}}
&  \frac{\sin{2 \theta} - 1}{\sqrt{2}} \\
\frac{\sin{2 \theta} - 1}{\sqrt{2}} &
- \frac{\cos{2 \theta}}{2} &  - \frac{\cos{2 \theta}}{2} \\
\frac{\sin{2 \theta} - 1}{\sqrt{2}} &
- \frac{\cos{2 \theta}}{2} &  - \frac{\cos{2 \theta}}{2} 
\end{array} \right)
\nonumber \\ &\pm&
\tilde{m} \frac{\varepsilon_{atm}}{4} \cos^2{2 \theta}
\left( \begin{array} {c@{\quad}c@{\quad}c@{\quad}c}
0 & 0 & 0  \\
0 &   1 & -1\\
0 & -1 &1
\end{array} \right) \nonumber \\ &+&
\tilde{m} \cos{2 \theta} \frac{\varepsilon_{sol}}{8}
\left( \begin{array} {c@{\quad}c@{\quad}c@{\quad}c}
0 & \sqrt{2} \sin{2 \theta} & \sqrt{2} \sin{2 \theta} \\
\sqrt{2} \sin{2 \theta} & - 2 \cos{2 \theta} 
& -2 \cos{2 \theta} \\ \sqrt{2} \sin{2 \theta} &
-2 \cos{2 \theta}  & -2 \cos{2 \theta} 
\end{array} \right),
\label{MB3thetapi}
\end{eqnarray}
Here we recover the C1 and C2 textures of 
Altarelli-Feruglio (for the bimaximal mixing case) and see 
how the LMA solution can be obtained adding a perturbation 
from these textures which are the only ones where the neutrino masses
in the limit $\delta m^2_{atm}=\delta m^2_{sol}=0$ 
doesn't coincide with $m_{ee}$ but can be much larger than $m_{ee}$. 
Therefore these textures are the most relevant ones 
for dark matter. In the limit of exact bimaximal mixing 
and $m_{ee}$ fixed this 
neutrino mass is infinite. However as Eqs.~(\ref{MB1thetapi}) 
and (\ref{MB3thetapi}) show,
from these two textures we can get the LMA 
solution (with $\theta < \pi/4$) from the exact bimaximal case 
considering $m_{ee}$ as a perturbation with respect 
to $\tilde{m}=m_{ee}/\cos{2 \theta}$. 
Because of this involved
nature of the solution, we could get a LMA solution in this way even if
the 11 element vanishes in the zero texture, which wouldn't be so  
easy to see starting directly from the texture zero as in 
the analysis of Ref. \cite{altardelanazione}.

\item \underline{{The inverse hierarchy case}}: 
We could have an inverse hierarchy solution, only if $m_{ee}$ 
turns out to be not much larger than the 
lower edge of the
range of Eq.~(\ref{meeexp}), i.e. $m_{ee}=0.05$~eV.
This is possible only with solutions $B_{2,4}$,
because as with solutions $A_{2,4}$, 
$\delta m^2_{atm}$ has to come with an opposite 
sign with respect to the term
in $m_{ee}^2$ in $m_3$, Eqs.~(\ref{m3B1})-(\ref{m3B3}). For 
these solutions $B_{2,4}$ an inverse hierarchy requires
$m^2_{ee} \sim \cos^2{2 \theta} \delta m^2_{atm}$. 
For the central values
$\delta m^2_{atm}=2.9 \cdot 10^{-3}$~eV$^2$ \cite{foglilisi} and 
$\tan^2{\theta}=0.37$ \cite{bahcallnov}, this requires 
for the LMA solution
$m_{ee}\sim 0.025$~eV.
Varying 
$\tan^2{\theta}$
within the range 0.2-0.8 \cite{bahcallnov} and 
$\delta m^2_{atm}$ within 
the range $(1.5-4.8) \cdot 10^{-3}$ eV$^2$ this requires: 
\begin{equation}
5 \cdot 10^{-3} \,\,\mbox{eV} \, <\, m_{ee}\, <\, 5 \cdot 10^{-2}\,\,\mbox{eV}
\,\,\, \mbox{(for 
solutions $B_{2,4}$ and LMA)}\label{meerange2}
\end{equation}
For the SMA solution since $\cos{2 \theta}$ is very close from unity the 
corresponding allowed range 
coincides with the one of Eq.~(\ref{meerange1}).
Note finally that if we develop the mass matrix around $\theta=0$ and
$\theta=\pi/4$, considering $m_3$ and
$\sqrt{\delta m^2_{sol}}$ as perturbations with respect to $\tilde{m}$ and
$\sqrt{\delta m^2_{atm}}$ we obtain the following zero textures:
\begin{equation}
\left( \begin{array} {c@{\quad}c@{\quad}c@{\quad}c}
1 & 0 & 0  \\
0 & -1/2 & -1/2 \\
0 & -1/2 & -1/2  
\end{array} \right), \,\,\,\,\,\mbox{and} \,\,\,\, 
\left( \begin{array} {c@{\quad}c@{\quad}c@{\quad}c}
0 & 1/\sqrt{2} & 1/\sqrt{2}  \\
1/\sqrt{2} & 0 & 0 \\
1/\sqrt{2} & 0 & 0  
\end{array} \right) ,
\end{equation}
respectively (that is to say the textures B1 of Ref.~\cite{altardelanazione}). 
\end{itemize}
\vspace{2mm}

\noindent
{\bf 7)} \underline{\bf Ambiguities of the perturbations}: In 
Eqs.~(\ref{MA1})-(\ref{MB3})
we have given the structure of the perturbations which has been determined
directly by
Eqs.~(\ref{Mtotal})-(\ref{m3B3}). They are 
the minimal perturbations which have to be added 
to give the atmospheric mass difference together with the solar angle and
solar mass difference. It must be noted, however, 
that these perturbations are 
not unique. For example in Eq.~(\ref{MA1}) we could perfectly 
add a perturbation
which is simply proportional to $\varepsilon_{atm}$ and the identity matrix.
To add this perturbation corresponds to a redefinition of the mass scale of the
zero texture by a small amount.
More generally there exist  
perturbations proportional to $\varepsilon_{sol}$ 
we could add which are not proportional to the zero texture
and which could leave the condition $m_2^2-m_1^2=\delta m^2_{sol}$
unchanged while changing the scale of the zero texture or the 
atmospheric mass difference by a negligible term
proportional to $\delta m^2_{sol}$.
To add such a perturbation 
leads to a mass matrix still fulfilling all the experimental constraints, 
but with another structure of perturbation. 
Note that the atmospheric mass 
difference is anyway far from being determined at the level of 
accuracy of the solar mass difference. 
This leads to
some freedom in the perturbations which exists also for perturbations
proportional to $\delta m^2_{atm}$. 
We leave this discussion to a subsequent publication. Here we gave just the 
"minimal" perturbations and we do not 
consider these extra perturbations 
which are not necessary to satisfy the experimental constraints.

Although the atmospheric neutrino mixing is allowed to be in the
range of $\sin^2 2 \theta_{atm} = 0.87 - 1$, we considered only the maximal 
mixing and not the entire range. This and the possible small deviations
from the constraint $U_{e3}=0$ may allow for few more
perturbations, which we disregarded here.

\vspace{2mm}

\noindent
{\bf 8)} \underline{\bf Summary}: In summary,
instead of starting
the analysis from the possible textures, we started with the
experimental inputs assuming in addition that $m_{ee}$
is larger than the neutrino mass differences. We then determined the neutrino
mass matrices and in the process got the possible textures. We
found that, for maximal atmospheric mixing and $U_{e3} = 0$
and for a given set of values of $m_{ee}$, $\delta m^2_{sol}$,
$\delta m^2_{atm}$ and $\theta$, there are 8 solutions for the mass 
matrix.
Among them the solutions $A_1$ to $A_4$ have the interesting property
to allow a decomposition which is just a zero texture depending only
on a single mass scale, plus a perturbation
depending only on $\delta m^2_{atm}$, plus a perturbation
depending only on $\delta m^2_{sol}$ and $\theta$.
There is also a set of 
solution $B_1$ to $B_4$ whose structure of perturbations is
more complicated but which allow neutrino masses larger than $m_{ee}$.
All solutions and especially $B_{1,2,3,4}$ lead naturally to 
neutrino masses relevant for dark matter. 
For solutions $B_{1,2,3,4}$, a lower $m_{ee}$ and a higher 
$\cos 2{\theta}$ is preferred by the dark matter fits. 
We observed also that for solutions $B_{1,2,3,4}$ there 
is a continuum of possible zero textures.
For each solution we displayed the allowed ranges for the mass eigenvalues 
together with the corresponding zero
texture and 
we showed how, in the degenerate neutrino mass scenario, 
the perturbations come up
to give the full mass matrix. 

%
\vspace{1.1cm}
\begin{center}{\large Acknowledgements}
\end{center}
We thank H.V. Klapdor-Kleingrothaus and U. Sarkar for several useful
discussions and their encouragements.
We also acknowledge useful comments from M. Raidal and J. Orloff. 
This work was
supported by the TMR, EC-contract No. ERBFMRX-CT980169
(Euro\-Da$\phi$ne).

%
\vspace{0.9cm}
\noindent
\underline{Last minute note}: in the process of submitting this manuscript,
some of
the results derived here turn out to 
have been presented independently in Refs.~\cite{glashow,strumia2}.
%
\vspace{0.9cm}

\end{document}